\documentclass[pra,onecolumn,a4paper,superscriptaddress,showpacs,showkeys]{revtex4}
\usepackage{graphicx}
\usepackage{amsmath}
\usepackage{amsfonts}
\usepackage{amssymb}

\normalsize
\begin{document}

\preprint{\begin{tabular}{r}
UWThPh-2003-15\\
July 2003\\
\end{tabular}}
\title{Berry phase in entangled systems: a proposed experiment\\
with single neutrons}

\author{Reinhold A. Bertlmann}
\author{Katharina Durstberger}
\email{kadu@ap.univie.ac.at}
\affiliation{Institute for Theoretical Physics, University of Vienna\\
Boltzmanngasse 5, 1090 Vienna, Austria}
\author{Yuji Hasegawa}
\affiliation{Atominstitut der \"Osterreichischen Universit\"aten\\Stadionallee 2, 1020
Vienna, Austria}
\author{Beatrix C. Hiesmayr}
\email{hiesmayr@ap.univie.ac.at}
\affiliation{Grup de F\'{i}sica Te\`{o}rica, Universitat
Aut\`{o}noma de Barcelona\\E-08193 Bellaterra, Spain}

\begin{abstract}
The influence of the geometric phase, in particular the Berry phase, on an entangled
spin-$\frac{1}{2}$ system is studied. We discuss in detail the case, where the geometric
phase is generated only by one part of the Hilbert space. We are able to cancel the
effects of the dynamical phase by using the ``spin-echo'' method. We analyze how the
Berry phase affects the Bell angles and the maximal violation of a Bell inequality.
Furthermore we suggest an experimental realization of our setup within neutron
interferometry.
\end{abstract}


\maketitle

\section{Introduction}

Entanglement is one of the most profound aspects of quantum mechanics (QM). It occurs in
quantum systems that consist of two (or more) parts which can be separated, or typically,
in systems whose observables belong to disjoint Hilbert spaces. Its deep meaning has been
already realized by Erwin Schr\"odinger in 1935 in his famous papers on ``The present
situation of quantum mechanics'' \cite{Schrodinger}. In Schr\"odinger's view the {\it
whole} system is in a definite state whereas the individual parts are not.

In the same year Einstein, Podolsky and Rosen \cite{EPR} recognized -- what nowadays is
called ``EPR paradox'' -- that QM exhibits very peculiar correlations between two
physically distant parts of the total system. It is possible to predict the outcome of
the measurement of one part by looking at the distant part.

About three decades later John S. Bell \cite{Bell1964} re-analyzed the ``EPR paradox''.
He discovered inequalities, commonly known as Bell inequalities (BI), that can be
violated by QM but have to be satisfied by (all) local realistic theories
\cite{BellBook}. A violation of a BI demonstrates the presence of entanglement and thus
according to Bell's Theorem the occurrence of nonlocal features in the quantum systems.
Generalizations of such inequalities serve as criterion for entanglement or separability
\cite{BNT}.

Experiments with photons (see, e.g., Ref.\cite{BertlmannZeilinger} for a review) confirm
QM with its nonlocality in an impressive way. In the last years there have also been
considerable activities to test entangled massive systems in particle physics
\cite{BertlmannGrimus2001,BertlmannHiesmayr2001,BertlmannDurstbergerHiesmayr2002,%
BramonNowakowski,BramonGarbarino,BramonGarbarinoHiesmayr2003_1,GenoveseNoveroPredazzi}.
Entanglement is the basis for quantum communication and quantum information (see, e.g.,
Ref.\cite{BertlmannZeilinger}) and it became an important issue of investigation
nowadays.

Geometric phases such as the Berry phase \cite{Berry1} play a considerable role in
physics and arise in a quantum system when its time evolution is cyclic and adiabatic.
Its deep geometric origin is given by a holonomy of the line bundle of the states where
the phase emerges from the integral of the connection (or curvature) of the bundle over
the parameter space \cite{Simon}. The quantum holonomy also appears for mixed states
\cite{Uhlmann1986,DabrowskiGrosse1990}. Generalizations to nonadiabatic evolutions
\cite{AharonovAnandan} and noncyclic and nonunitary settings
\cite{Pancharatnam,SamuelBhandari} do exist as well as extensions to off-diagonal
geometric phases \cite{ManiniPistolesi,FilippSjoqvist}. The application of geometric
phases in quantum computation has been suggested by several authors
\cite{ZanardiRasetti,JonesVedralEkertCastagnoli,DuanCiracZoller}. Experimentally,
geometric phases have been tested in various cases, e.g., with photons
\cite{KwiatChiao,ChiaoWu,TomitaChiao}, with neutrons
\cite{HasegawaZawiskyRauchIoffe,HasegawaLoidlBadurekBaronManiniPistolesiRauch} and with
atoms \cite{WebbGodunSummy}.

Whereas the geometric phase in a single particle system is already studied very well,
both theoretically and experimentally, its effect on entangled quantum systems is less
known. However, there is increasing interest to combine both quantum phenomena, the
geometric phase and the entanglement of a system
\cite{Sjoqvist2000,TongKwekOh2003,MilmanMosseri2003,TongSjoqvistKwekOhEricsson}.

In this article we are studying the influence of the Berry phase on the entanglement of a
spin-$\frac{1}{2}$ system by considering a BI. The Berry phase is generated by
implementing an adiabatically rotating magnetic field into one of the paths of the
particles. Our goal is to propose an explicit experimental setup which eliminates the
dynamical phase, which would spoil the geometric effect, so that we are sensitive just to
the geometric phase. We can achieve this within neutron interferometry
\cite{BadurekRauchZeilinger,RauchWerner} which is an almost ideal tool to investigate the
evolution of a spin-$\frac{1}{2}$ system. In particular, when using a polarized beam
\cite{BadurekRauchSummhammer} we have entanglement between different degrees of freedom,
i.e., the spin and the path of the neutron. In this case it is physically rather
noncontextuality than locality which is tested experimentally
\cite{BasuBandyopadhyayKarHome,HasegawaLoidlBadurekBaronRauch}.

Noncontextuality means that the value of an observable does not depend on the
experimental context; the measurement of the observable must yield the value independent
of other simultaneous measurement. The question is whether the properties of individual
parts of a quantum system (or ensemble) do have definite or predetermined values before
the act of measurement -- a main hypothesis in hidden variable theories. The ``no-go
theorem'' of Bell-Kochen-Specker \cite{Bell1966,KochenSpecker} states that noncontextual
theories are incompatible with QM. More precisely, it is in general impossible to assign
to an individual quantum system a definite value for each set of observables (see, e.g.,
Refs.\cite{Mermin1993,SimonZukowskiWeinfurterZeilinger}).

In our case the observables, which belong to mutually disjoint Hilbert spaces, are the
spin and the path of the neutron in the interferometer and we use a BI containing these
observables to test noncontextual hidden variable theories versus QM
\cite{BasuBandyopadhyayKarHome}.

\section{The Berry phase for spin-$\frac{1}{2}$ particles}\label{magneticfield}

We concentrate on the spin-$\frac{1}{2}$ system where it is rather simple to implement a
geometrical phase and we use Berry's \cite {Berry1} construction for the system evolving
adiabatically and cyclically in time.

The scenario is as follows. The particle, without loss of generality moving in
$y$-direction, couples to a time dependent magnetic field $\vec{B}(t)$ with unit vector
$\vec{n}(\vartheta; t)$ and constant norm $B = |\vec{B}(t)|$. Field $\vec{B}(t)$ rotates
adiabatically with an angular velocity $\omega_0$ around the $z$-axis under an angle
$\vartheta$ (for proper adiabaticity conditions, see Ref.\cite{AguiarNemesPeixoto}). The
interaction is described by the Hamiltonian
\begin{equation}\label{Hamiltonian}
H(t)=\frac{\mu}{2} \vec{B}(t) \vec{\sigma} \; ,
\end{equation}
where the coupling constant is given by $\mu=g\mu_B$, the Land\'e factor $g$ times the
Bohr magneton $\mu_B=\frac{1}{2}\frac{e}{m}\hbar$.

Consequently, the instantaneous eigenstates of the spin-operator in direction
$\vec{n}(\vartheta ; t)$ -- thus of Hamiltonian (\ref{Hamiltonian}) -- expanded in the
$\sigma_z$-basis are given by
\begin{equation}\label{basis-transformation}
\begin{split}
    |\Uparrow_n; t\rangle&=\cos\frac{\vartheta}{2}\; |\Uparrow_z\rangle+
    \sin\frac{\vartheta}{2}\;e^{i\omega_0 t}\;|\Downarrow_z\rangle\\
    |\Downarrow_n; t\rangle&=-\sin\frac{\vartheta}{2}\; |\Uparrow_z\rangle+
    \cos\frac{\vartheta}{2}\;e^{i\omega_0 t}\;|\Downarrow_z\rangle \; .
\end{split}
\end{equation}
The corresponding time independent energy levels are
\begin{equation}
    E_{\pm} = \pm \frac{\mu B}{2} = \pm \hbar \omega_1 \; ,
\end{equation}
where $\omega_1:=\frac{E_+-E_-}{2\hbar}=\frac{\mu B}{2\hbar}$ denotes the energy difference
of spin $\Uparrow_n$ and spin $\Downarrow_n$ and represents the characteristic frequency of
the system. Let us consider an adiabatic (which means $\frac{\omega_0}{\omega_1}\ll1$) and
cyclic time evolution for the period $\tau=\frac{2\pi}{\omega_0}$ of these eigenstates. Then
each eigenstate picks up a phase factor that can be split into a geometrical and dynamical
part of the following form
\begin{equation}\label{phases}
\begin{split}
    \lvert\Uparrow_n; t=0\rangle \longrightarrow\lvert \Uparrow_n; t=\tau\rangle
    &=e^{i\gamma_+(\vartheta)} e^{i\theta_+}\lvert \Uparrow_n; t=0\rangle \\
    \lvert\Downarrow_n; t=0\rangle \longrightarrow\lvert \Downarrow_n; t=\tau\rangle
    &= e^{i\gamma_-(\vartheta)} e^{i\theta_-}\lvert \Downarrow_n; t=0\rangle \; ,
\end{split}
\end{equation}
with
\begin{align}\label{geom.phase}
    &\gamma_+(\vartheta) = -\pi(1-\cos\vartheta)&
    &\gamma_-(\vartheta) = -\pi(1+\cos\vartheta)=-\gamma_+(\vartheta) - 2\pi \\
    &\theta_+ = -\frac{1}{\hbar}E_+\tau= -2\pi\frac{\omega_1}{\omega_0}&
    &\theta_- = +\frac{1}{\hbar}E_-\tau= +2\pi\frac{\omega_1}{\omega_0}= - \theta_+ \; .
\end{align}
Symbol $\gamma_{\pm}$ denotes the Berry phase which is precisely half of the solid angle
$\frac{1}{2} \Omega$ swept out by the magnetic field during the rotation and
$\theta_{\pm}$ is the familiar dynamical phase.

Now we are going to eliminate the dynamical effect which would dominate the geometrical
one, by using the so called ``spin-echo'' method. First the propagating particle is
subjected to the rotating magnetic field in the direction $\vec{n}(\vartheta)$ for one
period and therefore picks up the phases given by Eq.(\ref{phases}). Afterwards the
particle passes another rotating field which points in direction
$-\vec{n}(\pi-\vartheta)$ again for one period. Then the states change according to
\begin{equation}
\begin{split}
    \lvert\Uparrow_n\rangle\equiv\lvert\Downarrow_{-n}\rangle&\longrightarrow
    e^{i\gamma_-(\pi-\vartheta)}e^{i\theta_-}\lvert\Downarrow_{-n}\rangle\equiv
    e^{i\gamma_+(\vartheta)} e^{i\theta_-}\lvert\Uparrow_n\rangle \\
    \lvert\Downarrow_n\rangle\equiv\lvert\Uparrow_{-n}\rangle&\longrightarrow
    e^{i\gamma_+(\pi-\vartheta)}e^{i\theta_+}\lvert\Uparrow_{-n}\rangle\equiv
    e^{i\gamma_-(\vartheta)} e^{i\theta_+}\lvert\Downarrow_n\rangle \; .
\end{split}
\end{equation}
Therefore we get the following net-effect after two rotation-periods
\begin{equation}
    \lvert\Uparrow_n\rangle\rightarrow e^{2i\gamma_+(\vartheta)} \lvert
    \Uparrow_n\rangle \qquad
    \lvert\Downarrow_n\rangle\rightarrow e^{2i\gamma_-(\vartheta)} \lvert
    \Downarrow_n\rangle \; ,
\end{equation}
or for two half-periods of rotation we have
\begin{equation}\label{pickup}
    \lvert\Uparrow_n\rangle\rightarrow e^{i\gamma_+(\vartheta)} \lvert
    \Uparrow_n\rangle \qquad
    \lvert\Downarrow_n\rangle\rightarrow e^{i\gamma_-(\vartheta)} \lvert
    \Downarrow_n\rangle \; ,
\end{equation}
where the dynamical effects totally disappear.

\section{The Berry phase and the entangled state}\label{Berry1Arm}

Let us consider an entangled state of two spin-$\frac{1}{2}$ particles, e.g., the
antisymmetric Bell singlet state $\Psi^{(-)}$. One of the particles (e.g., the left side
moving particle) interacts twice with the adiabatically rotating magnetic fields as
described in Sect.\ref{magneticfield}. Thus only one subspace of the Hilbert space is
influenced by the phases.

To locate the Berry phase we decompose the initial Bell singlet state into the eigenstates
of the interaction Hamiltonian
\begin{eqnarray}\label{initialstate-in-n-bases}
    \lvert\Psi(t=0)\rangle=\lvert\Psi^{(-)}\rangle=\frac{1}{\sqrt{2}}\lbrace
    \lvert\Uparrow_n\rangle_l\otimes\lvert\Downarrow_n\rangle_r-
    \lvert\Downarrow_n\rangle_l\otimes\lvert\Uparrow_n\rangle_r\rbrace \; .
\end{eqnarray}
According to our ``spin-echo'' construction, after one cycle, the state
(\ref{initialstate-in-n-bases}) picks up precisely the geometric phase (\ref{pickup})
\begin{eqnarray}
    \lvert\Psi(t=\tau)\rangle=\frac{1}{\sqrt{2}}\lbrace
    e^{i \gamma_+}\lvert\Uparrow_n\rangle_l\otimes\lvert\Downarrow_n\rangle_r
    - e^{i \gamma_-}\lvert\Downarrow_n\rangle_l\otimes\lvert\Uparrow_n\rangle_r
    \rbrace \; ,
\end{eqnarray}
which can be rewritten by neglecting an overall phase factor (from now on
$\gamma_+\equiv\gamma$)
\begin{eqnarray}\label{endzustand}
    \lvert\Psi(t=\tau)\rangle=\frac{1}{\sqrt{2}}
    \lbrace\lvert\Uparrow_n\rangle_l\otimes\lvert\Downarrow_n\rangle_r
    - e^{-2i\gamma}\lvert\Downarrow_n\rangle_l\otimes\lvert\Uparrow_n\rangle_r
    \rbrace \; .
\end{eqnarray}
Increasing the Berry phase $\lvert\gamma\rvert :
0\rightarrow\frac{\pi}{2}\rightarrow\pi$, which we achieve by varying the magnetic field
angle $\vartheta : 0 \rightarrow 60^{\circ}\rightarrow 90^{\circ}$, we move continuously
from the antisymmetric Bell singlet state $\Psi^{(-)}$ to the symmetric Bell state
$\Psi^{(+)}$ and back to $\Psi^{(-)}$.

\begin{figure}[ht]
\centering
     \includegraphics[width=15cm,height=10cm,keepaspectratio=true]{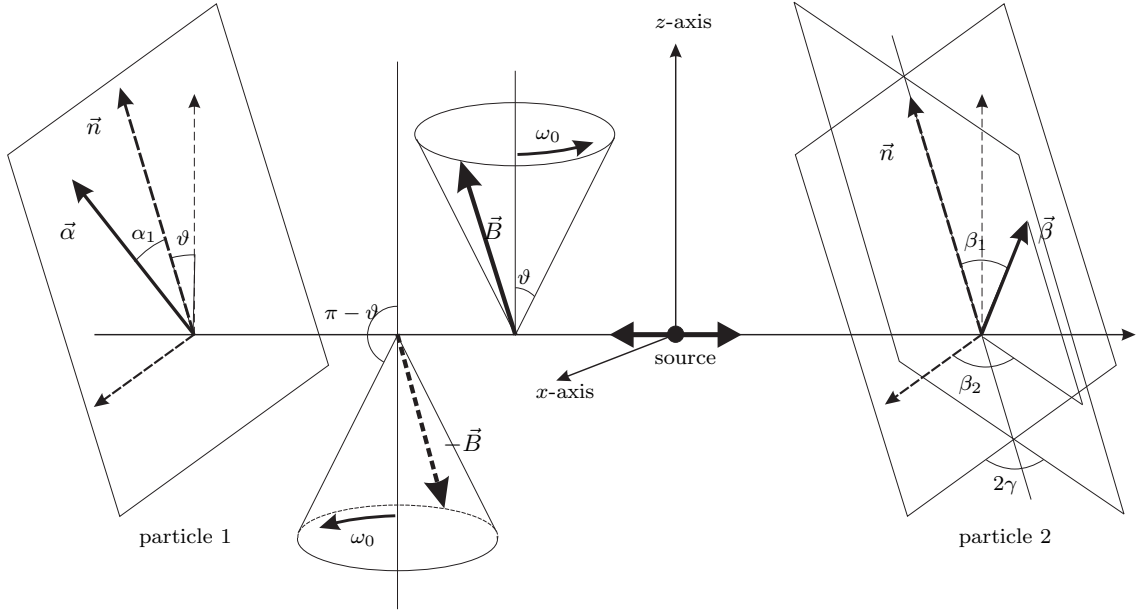}
    \begin{picture}(1,1)(0,0)
    \put(-250,140){\small{$\vec{B}$}}
    \put(-265,60){\small{$-\vec{B}$}}
    \put(-230,178){\footnotesize{$\omega_0$}}
    \put(-300,25){\footnotesize{$\omega_0$}}
    \put(-237,123){\footnotesize{$\vartheta$}}
    \put(-310,110){\footnotesize{$\pi-\vartheta$}}
    \put(-410,140){\small{$\vec\alpha$}}
    \put(-40,140){\small{$\vec\beta$}}
    \put(-400,180){\small{$\vec n$}}
    \put(-100,170){\small{$\vec n$}}
    \put(-383,140){\footnotesize{$\alpha_1$}}
    \put(-68,137){\footnotesize{$\beta_1$}}
    \put(-70,83){\footnotesize{$\beta_2$}}
    \put(-57,45){\footnotesize{$2\gamma$}}
    \put(-366,138){\footnotesize{$\vartheta$}}
    \put(-380,25){\footnotesize{particle 1}}
    \put(-70,25){\footnotesize{particle 2}}
    \put(-185,220){\footnotesize{$z$-axis}}
    \put(-230,80){\footnotesize{$x$-axis}}
    \put(-185,93){\footnotesize{source}}
  \end{picture}
     \caption{\footnotesize{Schematic view of the setup. The vector $\vec n$ denotes the
     quantization direction, $\vec\alpha$ and $\vec\beta$ are the measurement
     directions which determine the measurement planes.}} \label{theor_setup}
\end{figure}

As in common Bell experiments we want to measure simultaneously the spin components of
the particles on the left and right side, see Fig.\ref{theor_setup}. First we define the
projection operator onto an up (+) and down (-) spin state along an arbitrary direction
$\vec\alpha$
\begin{equation}
        P_{\pm}(\vec\alpha)=\lvert\pm\vec\alpha\rangle\langle\pm\vec\alpha\rvert
\end{equation}
with
\begin{equation}
\begin{split}
    \lvert+\vec\alpha\rangle &=
    \cos\frac{\alpha_1}{2}\lvert\Uparrow_n\rangle+
    \sin\frac{\alpha_1}{2}e^{i\alpha_2}\lvert\Downarrow_n\rangle\\
    \lvert-\vec\alpha\rangle &=
    -\sin\frac{\alpha_1}{2}\lvert\Uparrow_n\rangle+
    \cos\frac{\alpha_1}{2}e^{i\alpha_2}\lvert\Downarrow_n\rangle\;,
\end{split}
\end{equation}
where $\alpha_1$ denotes the polar angle measured from the $\vec n$-direction and
$\alpha_2$ the azimuthal angle. Then we calculate the joint probability for finding
spin-up on the left side under an angle $\vec\alpha$ from the quantization axis $\vec{n}$
and spin-up (or spin-down) on the right side under an angle $\vec\beta$
\begin{equation}\label{spinupspinup}
\begin{split}
    P(\vec\alpha \Uparrow_n,\vec\beta \Uparrow_n)
    &= \;
    \langle\Psi(t=\tau)\rvert P_+^l(\vec\alpha)\otimes P_+^r(\vec\beta)
    \lvert\Psi(t=\tau)\rangle\\
    &= \; \frac{1}{4}\bigl(1-\cos\alpha_1\cos\beta_1-
    \cos(\alpha_2-\beta_2+2\gamma)\sin\alpha_1\sin\beta_1\bigr)\\
    &\rightarrow \; \frac{1}{4}\bigl(1-\cos(\alpha_1-\beta_1)\bigr)
    \quad\textrm{for}\quad (\alpha_2-\beta_2)\rightarrow -2\gamma \; ,
\end{split}
\end{equation}
\begin{equation}\label{spinupspindown}
\begin{split}
    P(\vec\alpha \Uparrow_n,\vec\beta \Downarrow_n)
    &= \;
    \langle\Psi(t=\tau)\rvert P_+^l(\vec\alpha)\otimes P_-^r(\vec\beta)
    \lvert\Psi(t=\tau)\rangle\\
    &= \; \frac{1}{4}\bigl(1+\cos\alpha_1\cos\beta_1+
    \cos(\alpha_2-\beta_2+2\gamma)\sin\alpha_1\sin\beta_1\bigr)\\
    &\rightarrow \; \frac{1}{4}\bigl(1+\cos(\alpha_1-\beta_1)\bigr)
    \quad\textrm{for}\quad (\alpha_2-\beta_2)\rightarrow -2\gamma\; .
\end{split}
\end{equation}
Introducing the observable
\begin{equation}
    A^l(\vec\alpha)=P_{+}^l(\vec\alpha)-P_{-}^l(\vec\alpha)\; ,
\end{equation}
and similarly $B^r(\vec\beta)$, we calculate the expectation value of the joint
measurement
\begin{equation}\label{expecationvalue}
\begin{split}
    E(\vec\alpha,\vec\beta)&=\langle\Psi(t=\tau)\rvert A^l(\vec\alpha)\otimes B^r(\vec\beta)
    \lvert\Psi(t=\tau)\rangle\\
    &=-\cos\alpha_1\cos\beta_1-\cos(\alpha_2-\beta_2+2\gamma)\sin\alpha_1\sin\beta_1\\
    &\rightarrow \; -\cos(\alpha_1-\beta_1)
    \quad\textrm{for}\quad (\alpha_2-\beta_2)\rightarrow -2\gamma \; .
\end{split}
\end{equation}
We observe that we always can compensate the effect of the Berry phase by simply changing
the difference of the azimuthal angles $(\alpha_2-\beta_2)\rightarrow -2\gamma$ of the
two measuring directions $\vec\alpha$ and $\vec\beta$ and regain the familiar expressions
without Berry phase.

To test experimentally the influence of a pure geometric phase in the entangled state we
have to vary only the opening angle $\vartheta$ of the magnetic field, i.e., the geometry
of the setup, which is related to the Berry phase by formula (\ref{geom.phase}). Then we
measure expectation value (\ref{expecationvalue}) with respect to $\gamma$ at certain
fixed directions $\vec\alpha$ and $\vec\beta$. By rotating the measurement planes by the
angle difference $(\alpha_2-\beta_2)=-2\gamma$ the geometric effect is balanced.
Experimentally we propose to test this feature within neutron interferometry, see
Sect.\ref{experiment}.

When considering a BI to test the local features of the states we find the following
behavior, which we want to illustrate by considering the CHSH-inequality (Clauser, Horne,
Shimony, Holt) \cite{ClauserHorneShimonyHolt}, an experimentally testable type of a BI,
\begin{equation}\label{CHSHinequality}
    S\leq 2 \; ,
\end{equation}
where the $S$-function is defined by
\begin{equation}
\begin{split}
    S(\vec\alpha,\vec\alpha',\vec\beta,\vec\beta';\gamma)=&
    \bigl\lvert E(\vec\alpha,\vec\beta)-E(\vec\alpha,\vec\beta')\bigr\rvert
    + \bigl\lvert E(\vec\alpha',\vec\beta)+E(\vec\alpha',\vec\beta')\bigr\rvert\\
     =&\Bigl\lvert
     -\sin\alpha_1\bigl(\cos(\alpha_2-\beta_2+2\gamma)\sin\beta_1
    -\cos(\alpha_2-\beta'_2+2\gamma)\sin\beta'_1\bigr)\\
    &\quad-\cos\alpha_1\bigl(\cos\beta_1-\cos\beta'_1\bigr) \Bigr\rvert\\
    +&\Bigl\lvert
    -\sin\alpha'_1\bigl(\cos(\alpha'_2-\beta_2+2\gamma)\sin\beta_1
    +\cos(\alpha'_2-\beta'_2+2\gamma)\sin\beta'_1\bigr)\\
    &\quad-\cos\alpha'_1\bigl(\cos\beta_1+\cos\beta'_1\bigr)
    \Bigr\rvert\;.
\end{split}
\end{equation}
Without loss of generality we can eliminate one angle by setting, e.g., $\vec\alpha=0$
($\alpha_1=\alpha_2=0$), which gives
\begin{equation}\label{S-function}
\begin{split}
    S(\vec\alpha',\vec\beta,\vec\beta';\gamma)=&
     \Bigl\lvert -\sin\alpha'_1
    \Bigl(\cos(\alpha'_2-\beta_2+2\gamma)\sin\beta_1
    +\cos(\alpha'_2-\beta'_2+2\gamma)\sin\beta'_1\Bigr)\\
    &-\cos\alpha'_1(\cos\beta_1+\cos\beta'_1)\Bigr\rvert+
    \Bigl\lvert-\cos\beta_1+\cos\beta'_1\Bigr\rvert\;.
\end{split}
\end{equation}
We always can reach as maximal value of S the standard value $2\sqrt{2}$. We keep the
polar angles $\alpha'_1$, $\beta_1$ and $\beta'_1$ constant at the Bell angles
$\alpha'_1=\frac{\pi}{2}$, $\beta_1=\frac{\pi}{4}$, $\beta'_1=\frac{3\pi}{4}$ and adjust
the azimuthal parts
\begin{equation}
    S(\alpha'_2,\beta_2,\beta'_2;\gamma)=
     \sqrt{2}
    +\Bigl\lvert
    -\frac{\sqrt{2}}{2}\Bigl(\cos(\alpha'_2-\beta_2+2\gamma)
    +\cos(\alpha'_2-\beta'_2+2\gamma)\Bigr)\Bigr\rvert\;.
\end{equation}
The maximum $2\sqrt{2}$ is reached for $\beta_2=\beta'_2$ and
$\alpha'_2-\beta'_2=-2\gamma$ (mod $\pi$). For convenience we can choose $\alpha'_2=0$.
The measurement planes on both sides enclose an angle of $2\gamma$, see
Fig.\ref{theor_setup}.

\begin{figure}[htbp]
  \centering
  \vspace{0.5cm}
    \includegraphics[width=6cm,height=6cm,keepaspectratio=true]{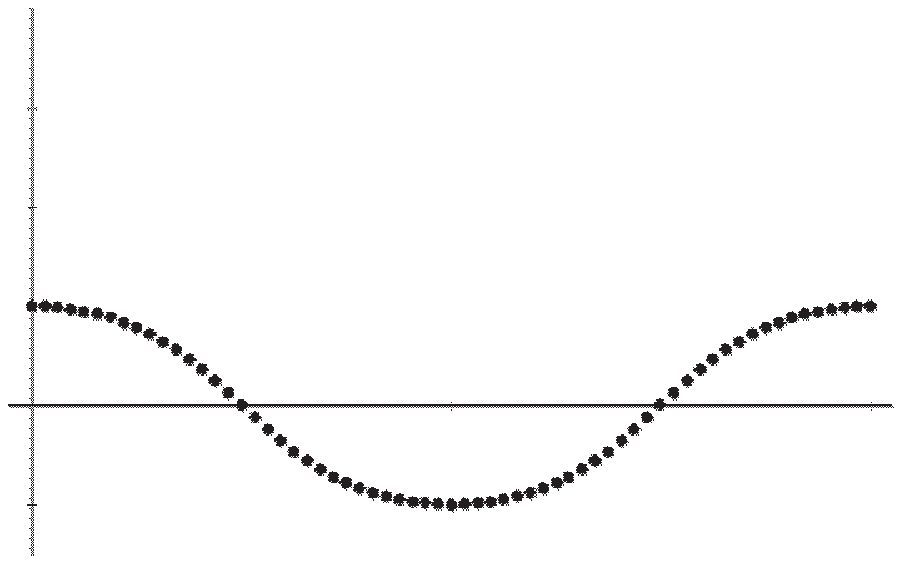}
    \begin{picture}(1,1)(0,0)
    \put(-180,110){ $\beta_1$-values}
    \put(-181,84){$\frac{3\pi}{4}$}
    \put(-180,65){$\frac{\pi}{2}$}
    \put(-180,46){$\frac{\pi}{4}$}
    \put(-185,8){$-\frac{\pi}{4}$}
    \put(-132,20){$\frac{\pi}{4}$}
    \put(-92,20){$\frac{\pi}{2}$}
    \put(-51,20){$\frac{3\pi}{4}$}
    \put(-11,20){$\pi$}
    \put(2,30){$\lvert\gamma\rvert$}
    \end{picture}
    \hspace{1cm}
    \includegraphics[width=6cm,height=6cm,keepaspectratio=true]{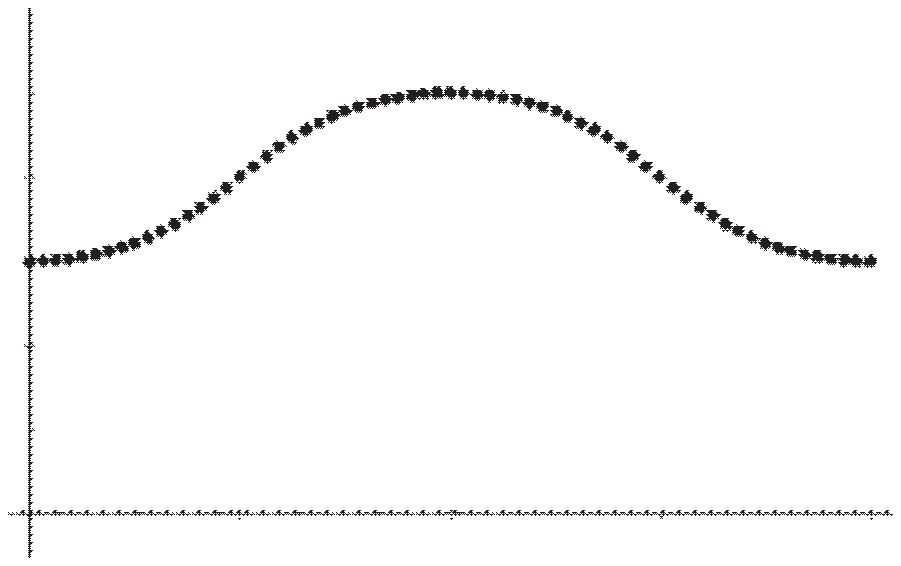}
    \begin{picture}(1,1)(0,0)
    \put(-180,110){ $\beta'_1$-values}
    \put(-181,87){$\frac{5\pi}{4}$}
    \put(-179,71){$\pi$}
    \put(-181,55){$\frac{3\pi}{4}$}
    \put(-180,39){$\frac{\pi}{2}$}
    \put(-180,23){$\frac{\pi}{4}$}
    \put(-132,0){$\frac{\pi}{4}$}
    \put(-92,0){$\frac{\pi}{2}$}
    \put(-54,0){$\frac{3\pi}{4}$}
    \put(-11,0){$\pi$}
    \put(2,10){$\lvert\gamma\rvert$}
    \end{picture}
    \caption{\footnotesize{The Bell angles $\beta_1$ and $\beta'_1$ with respect
    to the Berry phase $\gamma$ for the case $f_1<0$ and $f_2<0$.}}
  \label{beta-gamma-values1}
\end{figure}

\begin{figure}[htbp]
  \centering
    \includegraphics[width=6cm,height=6cm,keepaspectratio=true]{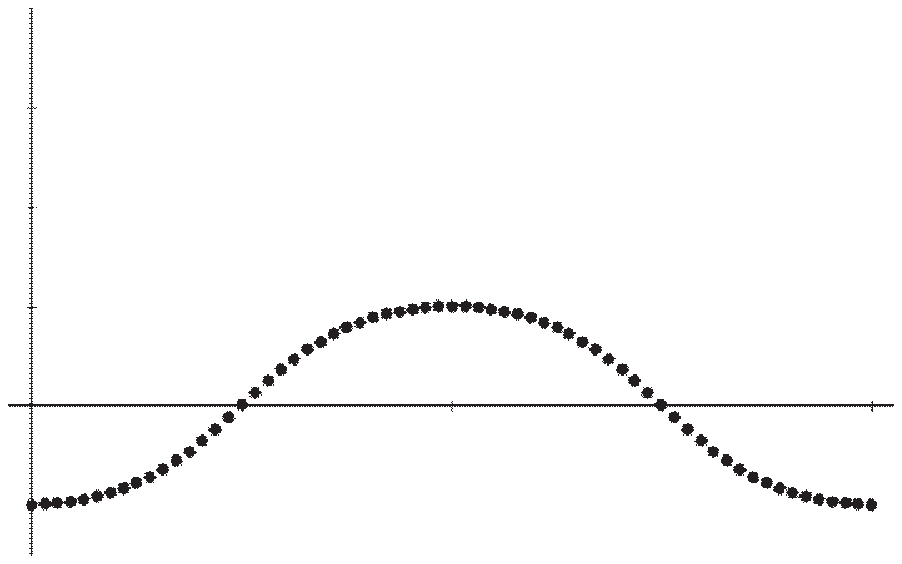}
    \begin{picture}(1,1)(0,0)
    \put(-180,110){ $\beta_1$-values}
    \put(-181,84){$\frac{3\pi}{4}$}
    \put(-180,65){$\frac{\pi}{2}$}
    \put(-180,46){$\frac{\pi}{4}$}
    \put(-185,8){$-\frac{\pi}{4}$}
    \put(-132,20){$\frac{\pi}{4}$}
    \put(-92,20){$\frac{\pi}{2}$}
    \put(-56,20){$\frac{3\pi}{4}$}
    \put(-11,20){$\pi$}
    \put(2,30){$\lvert\gamma\rvert$}
    \end{picture}
    \hspace{1cm}
    \includegraphics[width=6cm,height=6cm,keepaspectratio=true]{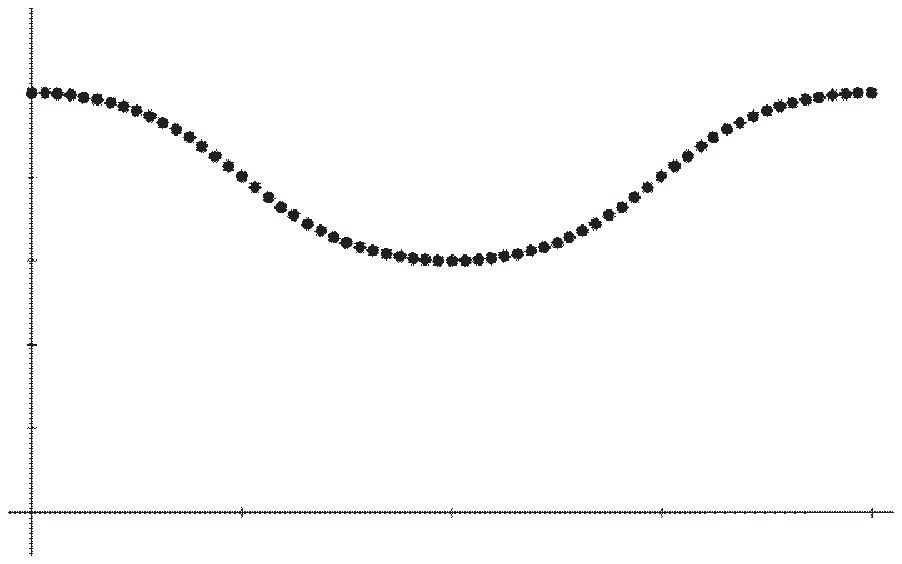}
    \begin{picture}(1,1)(0,0)
    \put(-180,110){ $\beta'_1$-values}
    \put(-181,87){$\frac{5\pi}{4}$}
    \put(-179,71){$\pi$}
    \put(-181,55){$\frac{3\pi}{4}$}
    \put(-180,39){$\frac{\pi}{2}$}
    \put(-180,23){$\frac{\pi}{4}$}
    \put(-132,0){$\frac{\pi}{4}$}
    \put(-92,0){$\frac{\pi}{2}$}
    \put(-54,0){$\frac{3\pi}{4}$}
    \put(-11,0){$\pi$}
    \put(2,10){$\lvert\gamma\rvert$}
    \end{picture}
    \caption{\footnotesize{The Bell angles $\beta_1$ and $\beta'_1$ with respect
    to the Berry phase $\gamma$ for the case $f_1<0$ and $f_2>0$.}}
  \label{beta-gamma-values2}
\end{figure}

On the other hand, keeping the azimuthal angles fixed, e.g.,
$\alpha_2=\alpha'_2=\beta_2=\beta'_2=0$, the polar Bell angles determined by the maximum
of the $S$-function change with respect to the Berry phase $\gamma$. By calculating the
derivatives (the extremum condition)
\begin{eqnarray}
    \frac{\partial S}{\partial\beta_1}  &=&
    -\sin\beta_1\mp\cos\alpha'_1\sin\beta_1\pm\cos(2\gamma)\sin\alpha'_1\cos\beta_1=0\nonumber\\
    \frac{\partial S}{\partial\beta'_1} &=&
    \sin\beta'_1\mp\cos\alpha'_1\sin\beta'_1\pm\cos(2\gamma)\sin\alpha'_1\cos\beta'_1=0\\
    \frac{\partial S}{\partial\alpha'_1} &=&
    \mp\sin\alpha'_1(\cos\beta_1+\cos\beta'_1)\pm\cos(2\gamma)\cos\alpha'_1
    (\sin\beta_1+\sin\beta'_1)=0\nonumber \;,
\end{eqnarray}
the solutions are given by ($\pm$ corresponds to the case either $f_1<0$ and $f_2<0$ or
to $f_1<0$ and $f_2>0$, when we denote $S= \lvert f_1\rvert + \lvert f_2\rvert$)
\begin{eqnarray}\label{Bell-angles}
\beta_1 &=& \pm\arctan(\cos(2\gamma))\nonumber\\
\beta'_1 &=& \pi - \beta_1\\
\alpha'_1 &=& \frac{\pi}{2}\nonumber \;,
\end{eqnarray}
and are plotted in Fig.\ref{beta-gamma-values1} and Fig.\ref{beta-gamma-values2} (of
course we may interchange $\beta_1 \leftrightarrow \beta'_1$) .

\begin{figure}[htbp]
  \centering
  \includegraphics[width=8cm,height=7cm,keepaspectratio=true]{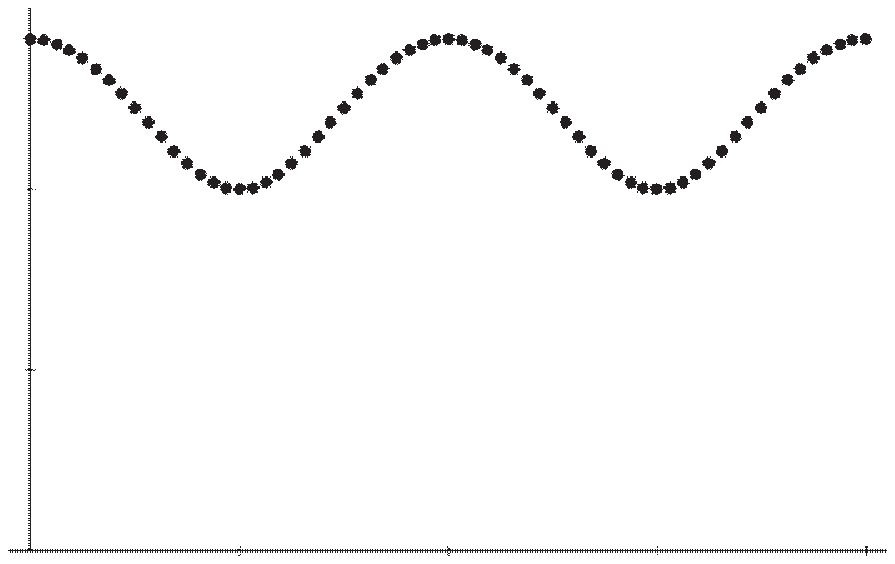}
  \begin{picture}(1,1)(0,0)
    \put(-250,148){maximal $S$-values}
    \put(9,10){$\lvert\gamma\rvert$}
    \put(-174,11){$\frac{\pi}{4}$}
    \put(-121,11){$\frac{\pi}{2}$}
    \put(-69,11){$\frac{3\pi}{4}$}
    \put(-13,11){$\pi$}
    \put(13,-10){$\vartheta$}
    \put(-225,-10){\small{$0^{\circ}$}}
    \put(-174,-10){\small{$41.4^{\circ}$}}
    \put(-121,-10){\small{$60^{\circ}$}}
    \put(-69,-10){\small{$75.5^{\circ}$}}
    \put(-13,-10){\small{$90^{\circ}$}}
    \put(-225,-25){$\Psi^{(-)}$}
    \put(-121,-25){$\Psi^{(+)}$}
    \put(-13,-25){$\Psi^{(-)}$}
    \put(-225,94){\line(1,0){215}}
    \put(-235,45){$1$}
    \put(-235,92){$2$}
    \put(-245,130){$2\sqrt{2}$}
  \end{picture}
  \vspace{0.5cm}
  \caption{\footnotesize{The maximum of the $S$-function (\ref{S-function}) with respect to
  the Berry phase $\gamma$ with the choice of zero azimuthal angles
  $\alpha'_2=\beta_2=\beta'_2=0$.}}
  \label{s_values}
\end{figure}

With these angles the $S$-function shows the behavior plotted in Fig.\ref{s_values}. We
see that the maximal $S$ decreases for $\gamma : 0\rightarrow\frac{\pi}{4}$ and touches
at $\gamma = \frac{\pi}{4}$ even the limit of the CHSH inequality $S=2$, where we are
unable to distinguish between QM and local realistic theories. It increases again to the
familiar value $S=2\sqrt{2}$ at $\gamma = \frac{\pi}{2}$, which corresponds to the Bell
state $\Psi^{(+)}$.


\section{Proposed neutron-experiment}\label{experiment}

Let us now consider how the predicted behavior of $S$ can be measured in practice. In our
polarized neutron interferometer experiment the wave function of each neutron is defined
over a tensor product of Hilbert spaces which describe the spatial and spin components of
the wave function and is entangled analogously to the two spin-$\frac{1}{2}$ system
(\ref{initialstate-in-n-bases})
\begin{equation}
    \left|{\Psi}\right\rangle={\frac{1}{\sqrt {2}}}
    \biggl\{\left|{\rm I}\right\rangle\otimes{\left|{s_{\rm I}}\right\rangle
    -\left|{\rm II}\right\rangle\otimes\left|{s_{\rm II}}\right\rangle }\biggr\}\;.
\end{equation}
The states $\left|{s_{\rm I}}\right\rangle$ and $\left|{s_{\rm II}}\right\rangle$ denote
the spin states of beam-I and -II, as well as $\left|{\rm I}\right\rangle$ and
$\left|{\rm II}\right\rangle$ the states in the two beam paths in the interferometer.

\begin{figure}[ht]
\centering
    \vspace{6mm}
    \includegraphics[width=10cm,height=10cm,keepaspectratio=true]{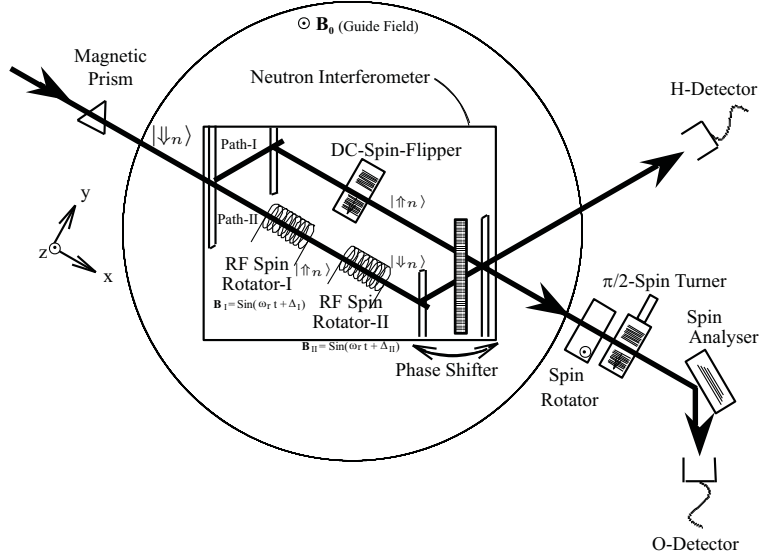}
  \begin{picture}(1,1)(0,0)
    \put(-233,155){{\footnotesize{$\lvert\Downarrow_n\rangle$}}}
    \put(-143,131){{\tiny{$\lvert\Uparrow_n\rangle$}}}
    \put(-179,105){{\tiny{$\lvert\Uparrow_n\rangle$}}}
    \put(-143,108){{\tiny{$\lvert\Downarrow_n\rangle$}}}
  \end{picture}
    \caption{\footnotesize{Schematic view of the experimental setup.}} \label{fig_setup}
\end{figure}

\begin{figure}[ht]
    \centering
    \includegraphics[width=3cm,height=4cm,keepaspectratio=true]{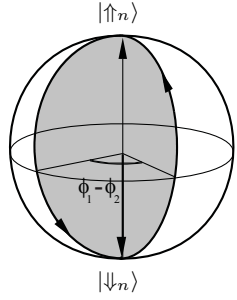}
    \begin{picture}(1,1)(0,0)
        \put(-55,105){{\footnotesize{$\lvert\Uparrow_n\rangle$}}}
        \put(-55,3){{\footnotesize{$\lvert\Downarrow_n\rangle$}}}
    \end{picture}
    \caption{\footnotesize{Schematic representation of the spinor evolution with the use of
    the Poincar\'{e} sphere.}} \label{fig_sphere}
\end{figure}

A schematic view of the experimental setup is shown in Fig.\ref{fig_setup}. In addition
to an auxiliary phase shifter, two radio-frequency (RF) spin-flippers
\cite{BadurekRauchTuppinger} are inserted into one beam path and a direct current (DC)
$\pi$-spin-flipper into the other path of the interferometer. The former two flippers
enable the neutron spinors to evolve along a particular curve inducing only a geometric
phase $\gamma_B$ without any dynamical component
\cite{WaghRakhechaSummhammer,AllmanKaiserWernerWaghRakhechaSummhammer}, see
Fig.\ref{fig_sphere}. The latter flipper produces the entangled state, like
$\left|{\Psi}(t=\tau)\right\rangle$ in Eq.(\ref{endzustand}). Thus, after the spinor
evolution the total wave function is represented by
\begin{equation}
\left|{\Psi(\gamma_B)}\right\rangle={\frac{1}{\sqrt {2}}} \biggl\{\left|{\rm
I}\right\rangle\otimes{\left|{\Uparrow_n}\right\rangle\rm - \rm {e}^{\mit i\rm \gamma_B }
\left|{\rm II}\right\rangle\otimes\left|{\Downarrow_n}\right\rangle }\biggr\}\;,
\end{equation}
with the geometrical phase
\begin{equation}
\gamma_B= \frac{1}{2}\Omega=\phi_1-\phi_2 .
\end{equation}

Our observables can be decomposed in the following form
\begin{equation}
    A^p(\chi)=P_+^p(\chi)-P_-^p(\chi)\quad\mbox{and}\quad
    B^s(\vec\delta)=P_+^s(\vec\delta)-P_-^s(\vec\delta)\;,
\end{equation}
where $P_{\pm}^p(\chi)$ and $P_{\pm}^s(\vec\delta)$ denote the projection operators onto
path and spin states, respectively
\begin{equation}
    P_{\pm}^p(\chi) =\lvert \pm p\rangle\langle \pm p \,\rvert\quad\mbox{and}\quad
    P_{\pm}^s(\vec\delta) =\lvert \pm \vec{n}_{\delta}\rangle
    \langle \pm \vec{n}_{\delta}\rvert\;,
\end{equation}
with
\begin{equation}
\begin{split}
    \lvert + p\rangle&=\cos\frac{\chi}{2}\lvert{\rm I}\rangle+
    \sin\frac{\chi}{2}\lvert{\rm II}\rangle \qquad\qquad
    \lvert +\vec{n}_{\delta}\rangle=\cos\frac{\delta_1}{2}\lvert\Uparrow_n\rangle+
    \sin\frac{\delta_1}{2}e^{i\delta_2}\lvert\Downarrow_n\rangle\\
    \lvert - p\rangle&=-\sin\frac{\chi}{2}\lvert{\rm I}\rangle+
    \cos\frac{\chi}{2}\lvert{\rm II}\rangle
    \qquad\quad
    \lvert -\vec{n}_{\delta}\rangle=-\sin\frac{\delta_1}{2}\lvert\Uparrow_n\rangle+
    \cos\frac{\delta_1}{2}e^{i\delta_2}\lvert\Downarrow_n\rangle\;.
\end{split}
\end{equation}

Once the state $\lvert\Psi(\gamma_B)\rangle$ is produced joint measurements on the path
and spin are performed by choosing the phase shift $\chi$ and the angle $\vec\delta$ of
the spinor analysis appropriately. Note that $\chi$ and $\vec\delta$ play the role of the
angles $\vec\alpha$ and $\vec\beta$ described before in Sect.\ref{Berry1Arm}.

Experimentally, the joint probabilities are given by the number of counts
\begin{equation}
\begin{split}
    N_{++}(\chi,\vec\delta)&=\langle\Psi(\gamma_B)\rvert P_+^p(\chi)\otimes
    P_+^s(\vec\delta)\lvert\Psi(\gamma_B)\rangle\\
    N_{+-}(\chi,\vec\delta)&=\langle\Psi(\gamma_B)\rvert P_+^p(\chi)\otimes
    P_+^s(\delta_1+\pi,\delta_2)\lvert\Psi(\gamma_B)\rangle=N_{++}(\chi,(\delta_1+\pi,\delta_2))\\
    N_{-+}(\chi,\vec\delta)&=\langle\Psi(\gamma_B)\rvert P_+^p(\chi+\pi)\otimes
    P_+^s(\vec\delta)\lvert\Psi(\gamma_B)\rangle=N_{++}(\chi+\pi,\vec\delta)\\
    N_{--}(\chi,\vec\delta)&=\langle\Psi(\gamma_B)\rvert P_+^p(\chi+\pi)\otimes
    P_+^s(\delta_1+\pi,\delta_2)\lvert\Psi(\gamma_B)\rangle=N_{++}(\chi+\pi,(\delta_1+\pi,\delta_2))\;.
\end{split}
\end{equation}
Then the expectation value
\begin{equation}
    E(\chi,\vec\delta)=\langle\Psi(\gamma_B)\rvert A^p(\chi)\otimes
    B^s(\vec\delta)\lvert\Psi(\gamma_B)\rangle
\end{equation}
is experimentally represented by
\begin{equation}\label{expectation_value_exp}
    E(\chi,\vec\delta)=
    \frac{N_{++}(\chi,\vec\delta)-N_{+-}(\chi,\vec\delta)-N_{-+}(\chi,\vec\delta)+N_{--}(\chi,\vec\delta)}
    {N_{++}(\chi,\vec\delta)+N_{+-}(\chi,\vec\delta)+N_{-+}(\chi,\vec\delta)+N_{--}(\chi,\vec\delta)}\;.
\end{equation}
The outcome of (\ref{expectation_value_exp}) coincides with formula
(\ref{expecationvalue}). Varying the opening angle of the magnetic field -- here it is
the relative phase of the RF between the two spin rotators -- one can demonstrate
experimentally the influence of the pure geometric phase of the entangled state on the
expectation value (\ref{expectation_value_exp}). Considering the CHSH inequality $S_{\rm
max}$ is achieved for $\chi\equiv\alpha_1=0$ ($\alpha_2=0$), which corresponds to the
choice of one path, and $\chi'\equiv\alpha'_1=\frac{\pi}{2}$ ($\alpha'_2=0$), which means
an equal superposition of the states $\lvert{\rm I}\rangle$ and $\lvert{\rm II}\rangle$,
whereas the angles $\vec\delta\equiv\vec\beta$ and $\vec\delta'\equiv\vec\beta'$ of the
spinor analysis have to be chosen accordingly.

\section{Summary and conclusion}

We have studied the influence of the Berry phase on an entangled spin-$\frac{1}{2}$
system, specifically the case where the Berry phase is generated by one subspace of the
system. Due to our special setup with external opposite rotating magnets, the
``spin-echo'' method, we are able to eliminate the dynamical phase such that only the
geometrical part remains. To analyze the effects of the Berry phase we consider the
expectation value of spin measurements and test the local feature via a CHSH inequality.

A phase -- like our geometrical one -- in a {\it pure} entangled system does {\it not}
change the amount of entanglement and therefore {\it not} the extent of nonlocality of
the system, which is determined by the maximal violation of a BI. Therefore such a phase
just affects the Bell angles in a definite way.

We always can achieve the familiar maximum value $2\sqrt{2}$ of the $S$-function by
rotating the Bell angles with respect to the Berry phase $\gamma$ by the azimuthal amount
$(\alpha_2-\beta_2)=-2\gamma$ as demonstrated in Fig.\ref{theor_setup}. This occurrence
of the maximal value is in accordance with Theorems of Horodecki \cite{Horodecki2_1996_1}
and Gisin \cite{Gisin1991}.

On the other hand, keeping the measurement planes fixed the polar Bell angles vary
according to formula (\ref{Bell-angles}) and the maximum of the $S$-function varies with
respect to the Berry phase $\gamma$ as shown in Fig.\ref{s_values}. It even touches the
boundary of the CHSH inequality at $\gamma = \frac{\pi}{4}$ making any distinction
between QM and local realistic theories obsolete for this setup. Other Bell inequalities
like Bell's original one \cite{Bell1964} show a similar behavior.

It is our free choice after all whether we rotate the measurement planes accordingly in
order to achieve the maximum value $S_{\rm{max}} = 2\sqrt{2}$ or keep them fixed at some
azimuthal angle and cope with smaller values of $S_{\rm{max}}$. In the second case we
show explicitly the dependence of $S_{\rm{max}}$ on the geometric phase.

Entanglement is a quantum property of states defined over a tensor product of Hilbert
spaces, no matter what kind of spaces they are. In this sense we certainly can entangle
the internal (spin) with the external (space) degrees of freedom of one and the same
particle. Then the physical interpretation of a BI, however, differs from the usual
nonlocality case and it is the more general concept of contextuality of the states, which
is tested, similarly to the Bell-Kochen-Specker Theorem
\cite{Bell1966,KochenSpecker,Mermin1993}. Noncontextuality here means that the value for
the observable spin of the neutron does not depend on the experimental context, i.e., on
the other co-measured observable, the path.

Noncontextuality is a rather restrictive demand for a theory, which is incompatible with
QM. We propose the experimental test with single particles -- the test of
noncontextuality versus QM -- to be performed within neutron interferometry which is an
excellent tool to study the properties of spin-$\frac{1}{2}$ systems. In our case we
study both the influence of the Berry phase on entanglement and on contextuality of the
states. The neutron experiment which can be easily carried out is in progress.

\begin{acknowledgments}

The authors want to thank Gerhard Ecker, Stefan Filipp, Heide Narnhofer, Helmut Rauch for
helpful discussions. This research has been supported by the EU project EURIDICE EEC-TMR
program HPRN-CT-2002-00311 and the FWF-project No. F1513 of the Austrian Science
Foundation.

\end{acknowledgments}

\bibliographystyle{prsty}
\bibliography{bibliographie}

\end{document}